\documentclass[prb,preprint,superscriptaddress,twocolumn,10pt]{revtex4-1} 
\usepackage{xcolor}
\usepackage{amsmath}  % needed for \tfrac, \bmatrix, etc.
\usepackage{amsfonts} % needed for bold Greek, Fraktur, and blackboard bold
\usepackage{graphicx} % needed for figures

\usepackage{placeins}

\usepackage{float}

\begin{document}

% Be sure to use the \title, \author, \affiliation, and \abstract macros
% to format your title page.  Don't use lower-level macros to  manually
% adjust the fonts and centering.

\title{A Curious Use of Extra Dimension in Classical Mechanics}
% In a long title you can use \\ to force a line break at a certain location.

\author{Trung V. Phan}
\email{tvphan@princeton.edu}
\affiliation{Department of Physics, Princeton University, Princeton, NJ 08544, USA}

\author{Anh Doan}
\affiliation{Department of Astronomy and Astrophysics, Pennsylvania State University, University Park, PA 16802, USA}

\begin{abstract}
Extra dimensions can be utilized to simplify problems in classical mechanics, offering new insights. Here we show a simple example of how the motion of a test particle under the influence of an 1D inverse-quadratic potential is equivalent to that of another test particle moving freely in 2D Euclidean space and 3D Minkowskian space.
\end{abstract}

\date{\today}

\maketitle

\section{Introduction}

Broadly speaking, physics questions are often defined by or complicated by their dimensions.  On the surface, there seems to be a trend that physical questions become more difficult as the dimensionality increases. For example, collisions in two dimensions are harder to deal with than in one dimension, because the velocity becomes a two-dimensional vector not a scalar. \cite{landau, morin_mechanics} Rigid body rotation in three dimensions is much more complicated than in two dimensions, since the angular velocity becomes a three-dimensional vector not a scalar. \cite{landau, morin_mechanics} In quantum mechanics, any potential-well in one dimension and two dimensions has at least one bound state, but that claim is no longer correct in three dimensions. \cite{yang_1D_2D, chadan_1D_2D} However, many findings in modern theoretical physics indicate that it is also possible to simplify problems by adding more dimensions. In the theory of general relativity, electromagnetism and gravity can be unified by adding an extra compact dimension. \cite{kaluza_unify, klein_unify} In condensed matter physics, quasicrystals can be treated as projections of a higher-dimensional lattice. \cite{kramer_periodic, levine_new, kalugin_6d} Finally, in string theory, a strongly interacting system can be more easily understood by considering a gravitational system in one more dimension \cite{hydrodynamic,condensed_matter} via gauge/string duality. \cite{ads_cft}

Though typically used in more advanced topics, the addition of extra dimensions can help to simplify problems in classical physics. For example, the electrostatic problem of finding the charge distribution on a thin conducting circular disk, can be easily solved by an orthogonal projection of the charge distribution of a conducting sphere onto an equatorial plane. \cite{thomson_paper} But, those are rare. In this note, we will concretely show how adding extra dimensions simplifies the classical mechanics problem of motion under the influence of an inverse-quadratic potential. To the best of our knowledge, this curious example has not been demonstrated elsewhere.

\section{From 1D to 2D Euclidean Space}

\begin{figure}[!htbp]
\centering
\includegraphics[width=0.45\textwidth]{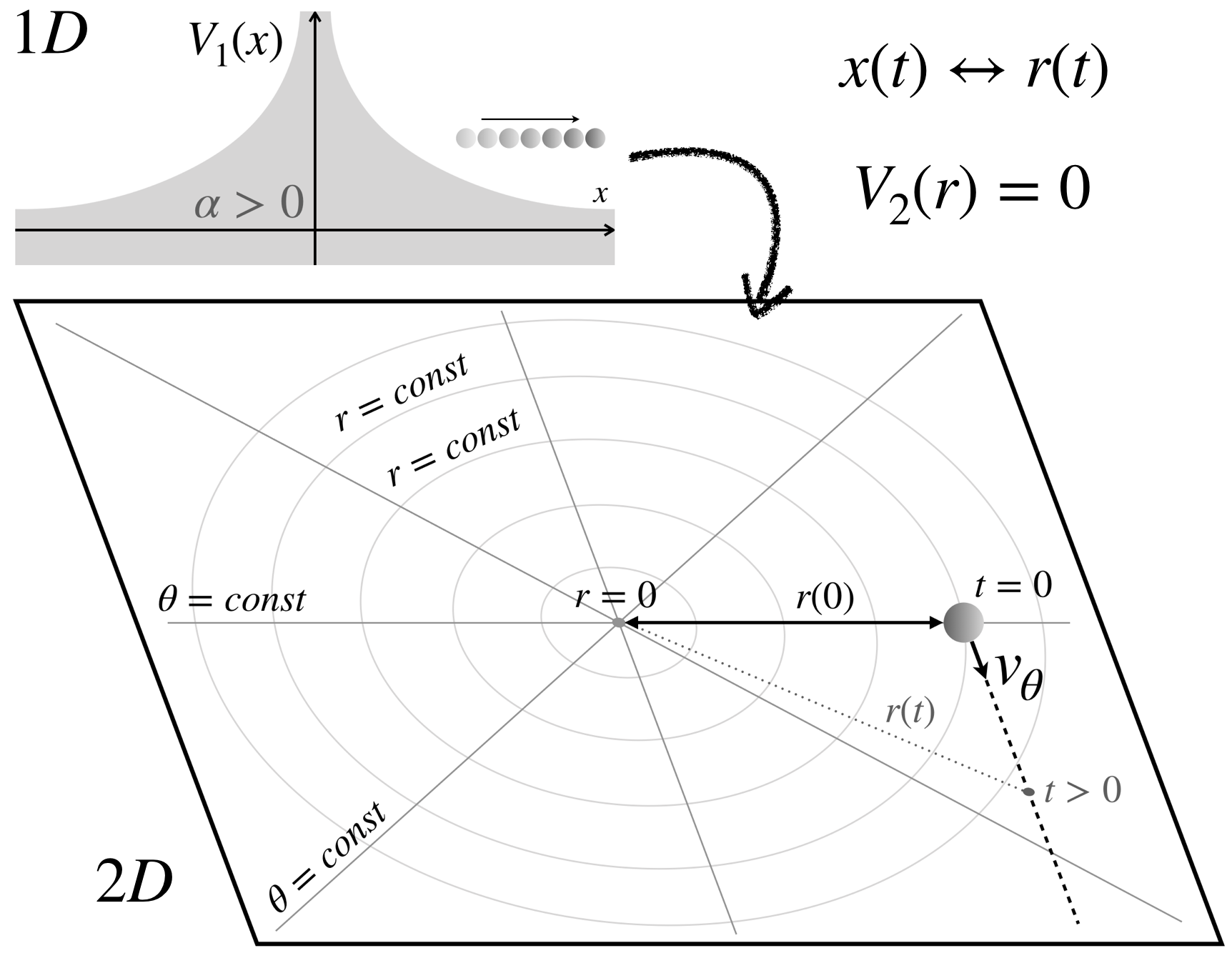}
\caption{The 1D/2D duality. A repulsive inverse-quadratic potential in an 1D space is dual to no potential in a 2D space.} 
\label{fig1}
\end{figure}

Consider a point particle of mass $m$ in one-dimensional space moving under the influence of an inverse-quadratic potential $V_{1}(x)=\alpha/x^2$. This potential appears in experimental atomic physics \cite{levy_1r2, francoise_1r2} and also is of many theoretical interests because it gives a scale-invariant Schrodinger's equation. \cite{nicholson_bound, coon_anomaly,quantum_1x2, quantum_alphax2} For now, let us focus on a repulsive potential with  $\alpha>0$. Initially, when $t=0$, the particle is at position $x_0$ with no velocity. The motion of the particle can be described by applying conservation of energy to arrive at the following integral:
\begin{equation}
\begin{split}
& m (dx/dt )^2/2 + V_{1}(x) = V_{1}(x_0) 
\\
\Rightarrow \ \ & dx/dt = \Big( 2\big(V_{1}(x_0) - V_{1}(x)\big)\big/m \Big)^{1/2}
\\
\Rightarrow \ \ & \int^{x(t)}_{x_0} dx\big( 2\alpha (x_0^{-2} - x^{-2} )/m \big)^{-1/2} = t \ \ . \ \ 
\end{split}
\end{equation}
However, doing this integration is non-trivial. The solution requires changing variables to $y=(x_0^2-x^2)^{1/2}$, at which point the integral becomes $\int dy y^{-1/2}$ up to a multiplication factor. We can get the equation of motion:
\begin{equation}
\begin{split}
& \Big( m x_0^2 \big(x^2(t) - x_0^2 \big)\big/2\alpha \Big)^{1/2} = t 
\\
\Rightarrow \ \ & x(t) = ( x_0^2 + 2\alpha t^2/m x_0^2 )^{1/2} \ \ . \ \ 
\end{split}
\label{answer}
\end{equation}
While this solution is tractable, there exists another way to describe the motion of the particle without the need of calculus. A ``magic'' from an extra dimension.

Consider a general central potential $V_2$ in a two-dimensional space. In the polar coordinates $\vec{r}=(r,\theta)$ where the origin is the center of the potential, we have rotational symmetry. The kinetic energy $K_\theta$ stored in the compact angular dimension depends on the angular momentum $p_\theta$ and the moment of inertia $m_\theta = mr^2$:
\begin{equation}
K_\theta = p_\theta^2/2m_\theta = p_\theta^2/2mr^2 = K_\theta(r)  \ \ . \ \ 
\end{equation}
The effective potential in the radial dimension \cite{landau, morin_mechanics} is just the sum of the central potential $V_2(r)$ and the kinetic energy $K_\theta(r)$. It should be noted that $K_\theta(r)$ and $V_1(x)$ are both inverse-quadratic functions.

\begin{figure*}[!th]
\centering
\includegraphics[width=0.8\textwidth]{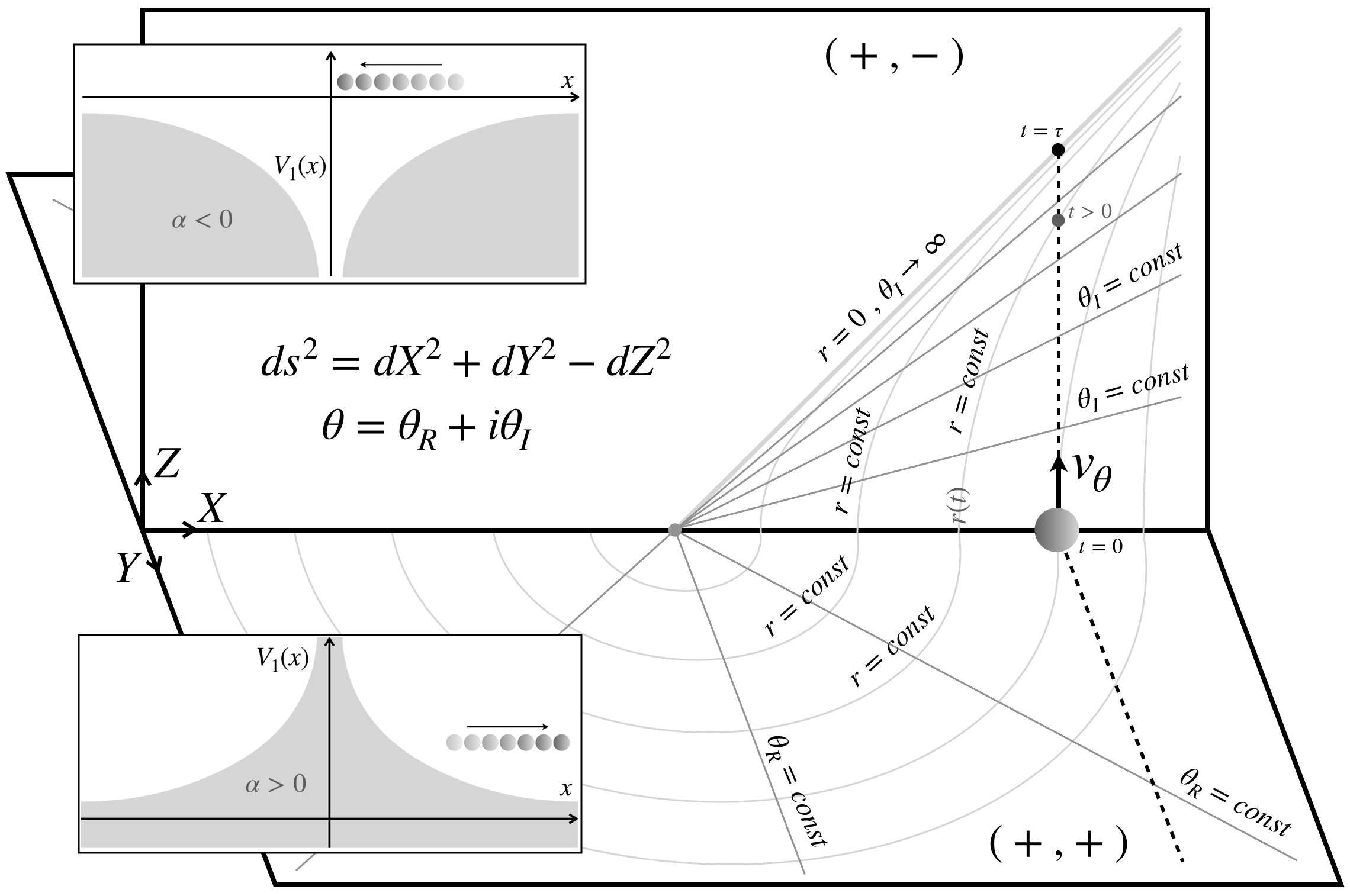}
\caption{The 1D/3D duality. An inverse-quadratic potential in an 1D space is dual to no potential in a 3D space with $(+,+,-)$ metric signature. Different signs of the potential correspond to different planes of motion: $(+,+)$ plane when the potential is repulsive and $(+,-)$ plane when the potential is attractive.} 
\label{fig2}
\end{figure*}

We note now an exact correspondence between this 2D scenario and the 1D problem considered above. The motion of the point particle in one-dimensional space under $V_1$ potential is dual to the radial motion of its counterpart moving freely in two-dimensional space (no potential $V_2=0$), given that the angular momentum is exactly $p_\theta = (2m\alpha)^{1/2}$:
\begin{equation} 
x(t)\Big|_{V_1 = \alpha/x^2} \ \ \leftrightarrow \ \ r(t)\Big|_{V_2 = 0, p_\theta = (2m\alpha)^{1/2}} \ \ . \ \ 
\end{equation}
In other words, with an extra angular dimension we can eliminate the potential.

Using $x(0)=x_0$ and $dx/dt(0)=0$, we have the corresponding radial position $r(0)=x_0$ and radial velocity $dr/dt(0) = 0$. The tangent velocity is given by:
\begin{equation} 
v_\theta \equiv rd\theta/dt(0) = p_\theta/mr(0) = (2\alpha/mx_0^2)^{1/2} \ \ . \ \ 
\end{equation}
See Fig. \ref{fig1} for the detail of this 1D/2D duality. We can arrive at the same answer \eqref{answer} with the Pythagorean theorem:
\begin{equation} 
r(t) = \big( r^2(0) + (v_\theta t)^2 \big)^{1/2} = ( x_0^2 + 2\alpha t^2/mx_0^2 )^{1/2} \ \ . \ \ 
\end{equation}
While we arrived at the same answer, the solution this time is purely geometric and does not involve any calculus.

\section{From 1D to 3D Minkowskian Space}

The tools developed here  can also be used for an arbitrary inverse-quadratic potential. However, it is more complicated and requires generalization to three dimensions. For an attractive potential, we can directly use $\alpha = -|\alpha|< 0$ to get the equation of motion $x(t)$ and also the lifetime $\tau$ until the point particle meets the singularity at position $x=0$:
\begin{equation} 
\begin{split}
& x(t)= \big( x_0^2 - 2|\alpha| t^2/mx_0^2 \big)^{1/2} \ \ , \ \ 
\\
&x(\tau) = 0 \ \ \Rightarrow \ \ \tau = \big(mx_0^4/2|\alpha|\big)^{1/2} \ \ . \ \ 
\end{split}
\end{equation}
However, as we re-examine the problem from a two-dimensional perspective as explained above, this indicates an imaginary value of angular momentum $p_\theta = i\big(2m|\alpha|\big)^{1/2}$. To generalize this extra-dimensional trick for all real values of $\alpha$, we need to complexify the angular dimension $\theta = \theta_R + i\theta_I$ (with $\theta_R$ and $\theta_I$ are real). Thus the corresponding space will be three-dimensional with $(+,+,-)$ metric signature. \cite{landau_vol2} Note that there are now two extra dimensions instead of one: while $\theta_R$ is a compact dimension, $\theta_I$ is an open one. The particle moves in the $(+,+)$ Euclidean plane when $\alpha>0$, and in the $(+,-)$ Minkowskian plane when $\alpha<0$. See Fig. \ref{fig2} for the detail of this 1D/3D duality.

\begin{acknowledgments}

This note comes from our final projects for MIT's course 8.223 in 2012 and 2013. We thank Duy V. Nguyen and the xPhO club for their support to share this finding to a wider audience. We thank Qiantan Hong and Leonid Levitov for examples in physics using extra dimensions. We also thank Oak Nelson and Todd Springer for many useful comments to make this note more approachable for general readers.

\end{acknowledgments}

\end{document}